\documentclass[aps,pra,epsf,twocolumn,showpacs,nofootinbib,amssymb,superscriptaddress]{revtex4-1}
\usepackage{graphicx}
\usepackage{amsmath}
\usepackage{dcolumn}
\usepackage{bm}
\usepackage{rotating}
\usepackage{soul}
\usepackage{amsmath}
\usepackage{epsfig}
\usepackage{epstopdf}
\begin{document}

\title{Chemical bond and entanglement of electrons in the hydrogen molecule}

\author{Nikos Iliopoulos}
\affiliation{Department of Physics, School of Natural
Sciences, University of Patras, Patras 265 04, Greece}
\author{Andreas F. Terzis}
\email{terzis@physics.upatras.gr}
\affiliation{Department of Physics, School of Natural
Sciences, University of Patras, Patras 265 04, Greece}

\date{\today}

\begin{abstract}
We theoretically  investigate the quantum correlations (in terms of concurrence of indistinguishable electrons)  in a prototype molecular system (hydrogen molecule). With the assistance of the standard approximations of the linear combination of atomic orbitals and the configuration interaction methods we describe the electronic wavefunction of  the ground state of the $H_{2}$ molecule. Moreover, we managed to find a rather simple analytic expression for the concurrence (the most used measure of quantum entanglement) of the two electrons when the molecule is in its lowest energy. We have found that concurrence does not really show any relation to the construction of the chemical bond. 
\end{abstract}


\maketitle

\section{Introduction}

During the last two decades a lot of work has been done  in the relatively new scientific field of quantum information and quantum computation \cite {nielsen}. In this new field of physics it became soon apparent that one of the most, if not the most,  important physical quantity -a measure of quantum correlations- is the quantum entanglement \cite{nielsen,austriakos}. Historically, entanglement is the first type among quantum correlations which studied intensively demonstrating the difference between classical and quantum mechanics.  Dated back to the Schr\"{o}dinger era, introduced by him to describe the correlations between two particles that interact and then separate, as in the EPR expertiment.  Actually, nowadays exist many measures of quantum entanglement. Entanglement of formation (EoF) or its computationally equivalent, concurrence (Con), is the most widespread among them \cite{wootters}. It is widely accepted  that EoF is a very reasonable and decent  measure of quantum entanglement having the advantage that, in most cases, it can be computed easily. However, it became soon apparent that there are other quantum correlations which cannot be included in  entanglement. Quantum discord \cite{{ollivier},{henderson}} is the most well-studied measure for the total quantum correlations. Quantum discord takes into account all possible sources of quantum correlations but its computation, requiring optimaztion procedures,  is much more complicated and time consuming compared to the corresponding computation of the Con. 

All previously mentioned measures are well-studied for distinguishable particles. However, they are not so widely used in indistinguishable particles. Especially for fermions (indistinguishable particles, having half-integer spin) obeying the Pauli exclusion principle, there are many difficulties in order to implement the measures used for distinguishable particles. Schliemann et al. \cite{schliemann2,schliemann1} have, recently, presented a corresponding measure of Con of fermionin pure states, taking into account the constraints due to the Pauli exclusion principle.  

In the present study we investigate the simplest (prototype) realistic two- electrons  molecular system, the hydrogen molecule which has been extensively  studied \cite{atkins} for almost a century. Due to the well known principle of minimum energy the chemical bond is traditionally described and specified in energy terms. Hence, we concentrate in the study of the ground state of the molecular system which is the most stable state, especially in the low temperature case where the thermal energy is not enough to excite the molecule. More specifically, in the present article, we investigate the possibility of any relationship between quantum  correlations and the chemical bond.   

This paper is organized as follows: In section II we present the basic theory of indistinguishable particles and we highlight the differences between indistinguishable and distinguishable ones.  In the same section, we find an analytical expression  for the Con of any two electrons being in a pure quantum state. In section III we apply this formula to the hydrogen molecule being in its ground state. Also, we apply the configuration interaction  (CI) method  in order to properly describe the ground state of the hydrogen molecule. In the same section we present and  analyze our results. Finally, we summarize and conclude in section IV.

\section{Theory}

To begin with, we report some basic properties of indinguishable particles, which are necessary in order to develop the quantum correlation identifiers. A general pure state of two fermions is given by the equation \cite{schliemann1,schliemann2} 
\begin{equation}
|w\rangle= \sum\limits_{i,j=1}^n w_{ij} f_{i}^\dagger f_{j}^\dagger |0\rangle
\label{eq1},
\end{equation}
$ f_{i}^\dagger, f_{j}^\dagger$ are the single particle creation operators for two corresponding subsystems acting on the vacuum state and $n$ is the dimensionality of the single-particle (one- electron) Hilbert space. Also, $w$ is an antisymmetric coefficient complex square matrix $(n \times n)$, with $w_{ij}=-w_{ji}$. The antisymmetric matrix fullfils the normalization condition
\begin{equation}
Tr(w^\ast w)=-1/2
\label{eq2}
\end{equation}

According to Ref. \cite{schliemann2} an equivalent to Schmidt decomposition \cite{austriakos} can be constructed in the case of fermions. Under a unitary transformation of $f_i =\sum_j U_{ij}f_{j}^\prime$, we take new fermionic operators as well as new coefficient matrix $w^\prime$. So, the pure fermionic state of Eq. (\ref{eq1}) takes the form
\begin{equation}
|w\rangle= \sum\limits_{i,j=1}^n w_{ij}^\prime {f_{i}^\prime}^\dagger {f_{j}^\prime}^\dagger |0\rangle
\label{eq3}
\end{equation}
where $w_{ij}^\prime = (U^\dagger w U^\ast)_{ij}$. This new matrix have a block diagonal form \cite{schliemann3}, containing $2 \times 2$ blocks of the type
\begin{equation}
\left[ \begin{array}{cc} 0 & z_k \\ -z_k & 0 \end{array} \right]
\label{eq4}
\end{equation}
Each block has eigevalue $z_k$. As a result the state of Eq. (\ref{eq1}) takes the diagonal form
\begin{equation}
|w\rangle= 2 \sum\limits_{k=1}^{\leq \frac{n}{2}}z_k  {f^\prime}^\dagger_{2k-1} {f^\prime}^\dagger_{2k} |0\rangle 
\label{eq5}
\end{equation}
This is the equivallent of Schmidt decomposition in indistinguishable particles \cite{austriakos}. Instead, now we have a sum of $2 \times 2$ blocks (Slater determinants) instead of sum of product states. We know in the case of distinguishable particles that if the Schmidt number is one then our state is a product state. Respectively, in order to have pure state in indistinguishable particles the Slater number, the equivalent of Schmidt number in this case, should be one too, i.e. our state  of eq. (\ref{eq5}) must contain only one Slater determinant. Otherwise it is an entangled state.  

Next, we give the definition of the entropy and the Con for the case of indistinguishable particles.

According to  Schliemann et al. \cite{schliemann3}, Con($C$)  for such a pure state of eq.  (\ref{eq1}) with $n=4$ is given by the equation
\begin{equation}
C(|w\rangle)=8|w_{12}w_{34}+w_{13}w_{42}+w_{14}w_{23}| 
\label{eq6},
\end{equation}
$w_{ij}$ are elements of the antisymmetric matrix $w$. As in the case of distinguishable particles, Con takes values from zero to one. If Con is zero then the fermionic state has no entanglement and it has fermionic slater rank one and could be represented by only one Slater determinant.

Now, in order to find an equation for the entropy of each subsystem we have to trace out the other. Due to the fact that fermions are indistinguishable particles the two parties have the same entropy. Also, the two fermion density matrix is given by $\rho_F=|w\rangle \langle w|$. According to R. Paskauskas and L. You \cite{you} the single particle density matrix is
\begin{equation}
\rho_{\nu \mu}^f=\frac{Tr(\rho_F f_{\mu}^\dagger f_{\nu})}{Tr(\rho_F \sum_{\mu}f_{\mu}^\dagger f_{\mu})}=2(\omega^\dagger \omega)_{\mu\nu}
\label{eq7}
\end{equation}
 Now, the normalization condition takes the form $ \sum\limits_{k=1}^{\leq n}|z_k|^2=1/4$. Finally, the von Neumann entropy is given by the equation
\begin{equation}
S_f=-Tr[\rho^f \log(\rho^f)]=-1-4\sum\limits_{k=1}^{\leq n}|z_k|^2 \log(|z_k|^2)
\label{eq8}
\end{equation}
This single particle entropy ranges from unit to $\log(n_E)$ and $n_E$ is the largest even number not larger than n.

Both Con and entropy are measures of the entanglement of our system. However, concurrence ranges from zero to unit, as in the case of separable particles, but entropy takes values $1\leq S_f \leq 2$ in contrast to the case of indistinguishable particles which ranges from zero to unity. This may have a simple interpetation. First of all, the von Neumann entropy is the measure of uncertainty that we have before performing a measurement. For example, if we have two separable particles which are in a Bell state, e.g. $\frac{1}{\sqrt{2}}(|0\rangle_A \otimes |0\rangle_B + |1\rangle_A \otimes |1\rangle_B)$, and we want to find the state of particle A then the von Neumann entropy is unit as the uncertainty about its state is maximum. In the case of indistinguishable particles, we cannot separate them, thus there is extra uncertainty because we do not know to which particle we perform the measurement. Nevertheless, this uncertainty does not affect the range of concurrence. 

From another point of view, the entanglement in the case of sepable particles is due to spatial coordinates or spin. For instance, the states $|0\rangle$ and $|1\rangle$ above could represent the possible states of spin (e.g. $|0\rangle =|\downarrow\rangle$ and $|1\rangle=|\uparrow\rangle$) or any other spatial separation. However, in order to have Con or entropy different than zero and one respectively, in the case of indistinguishable particles, the two particles should be entangled for spin and spatial coordinates. A fine example is given by Eckert et al. \cite{schliemann1}.  

In the hydrogen molecule there are two nuclei (protons) as well as two electrons occupying the region outside the protons. As a result each electron could be in nucleus A or B and has spin $|\uparrow\rangle$ or $|\downarrow\rangle$. Consequently, in our case the single-particle (one- electron) Hilbert space is four-dimensional and this resulting in a six-dimensional two-particle (two- electrons) Hilbert space. For this reason, from now on, we will deal with the pure state of eq. (\ref{eq1}) for $n=4$.

\section{Entanglement in hydrogen molecule}

Now, we turn our attention to the hydrogen molecule. As we said above, we have two electrons which wander around the two nucleus, A and B. Each electron could have spin +1/2 ($|\uparrow\rangle$) or -1/2 ($|\downarrow\rangle$). As a result the single-particle Hilbert space is four dimensional as it was reported above ($|A\rangle |\uparrow\rangle =|1\rangle$, $|A\rangle |\downarrow\rangle =|2\rangle$, $|B\rangle |\uparrow\rangle =|3\rangle$, $|B\rangle |\downarrow\rangle =|4\rangle$).

The Hamiltonian of our system is given by the equation (see for example, the classic book by  P. Atkins and R. Friedman \cite{atkins})
\begin{equation}
H=H_1 + H_2 + \frac{e^2}{4\pi \epsilon_0 R}+ \frac{e^2}{4\pi \epsilon_0 r_{12}}
\label{eq9}
\end{equation}
and $H_i=-\frac{\hbar^2}{2m_e}\nabla_i -\frac{e^2}{4\pi \epsilon_0 r_{iA}}-\frac{e^2}{4\pi \epsilon_0 r_{iB}}$, with $i=1,2$.

Here, R is the distance between the nuclei and $r_{12}$ the distance between the electrons. In addition, the two final terms represent the repulsive interaction between the nuclei and electrons respectively. 

As electrons are fermions, their wavefunction must be antisymmetric in order to obey in the Pauli exclusion principle. With the assistance of the linear combination of atomic orbitals (LCAO) method \cite{atkins} we conclude that the only possible wavefunctions of electrons are

\begin{subequations}
\begin{align}
\Psi_1=\psi_+ \psi_+ \sigma_-{(1,2)}\label{eq10a}\\
\Psi_2=\psi_- \psi_- \sigma_-{(1,2)}\label{eq10b}
\end{align}
\end{subequations}

$\psi_+=c_+ (\phi_A +\phi_B)$, $\psi_-=c_- (\phi_A -\phi_B)$ and $\sigma_{(1,2)}=\frac{1}{\sqrt{2}}( |\uparrow\rangle_1  |\downarrow\rangle_2 -  |\downarrow\rangle_1  |\uparrow\rangle_2)$. Here, $\phi_A$ and $\phi_B$ are the atomic orbitals of the hydrogen atom on ground state and $c_{\pm}$ are normaliztion factors which are $c_{\pm}=\frac{1}{\sqrt{2(1 \pm S)}}$, where $S$ is the overlap integral (see the appendix). 

\begin{figure}[htbp!]
\centering
\includegraphics*[width=9 cm]{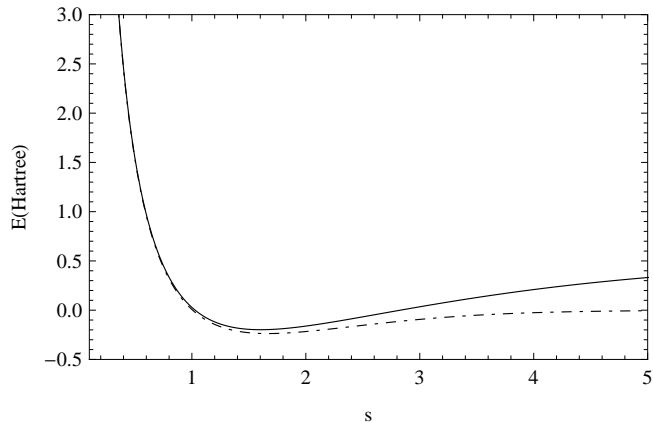}
\caption{Plots of the average energy of the $H_{2}$ molecule (actually the energy difference of the H molecule minus the energy of two separated H atom) in Hartree units (i.e. devided by the constant $hcR_H$) as a function of the internuclear distance ($s=R/\alpha_0$, where $\alpha_0$ is the Bohr radius). The solid curve corresponds to the energy (eq.(\ref{eq11})) of the molecule which is computed for a pure state described by eq.(\ref{eq10a}). The dashed-dotted curve is the energy estimated for pure state described by the linear combination  (CI method) of eq.(\ref{eq12}) }
\label{fig1}
\end{figure}
Using the wavefunction, $\Psi_1$, of equation (\ref{eq10a}) the hydrogen molecule has an average energy given by the expression
\begin{equation}
E_{1}=2E_{1s}+\frac{j_0}{R}-\frac{2j'+2k'}{1+S}+\frac{j+2k+m+4l}{2(1+S)^2}
\label{eq11}
\end{equation}
where $E_{1s}$ is the energy of the H atom in the ground state
Its plot as a function of the distance between the two nucleus is shows in Fig.(1) (solid curve).  Then, if we find the average energy of the  wavefunction given by the equation (\ref{eq10b}), we find that these is not an absolute minimum, and the energy is a decreasing function of the distance between the two protons and finally taking a plateau value equal to the plateau value of the energy of wavefunction of equation (\ref{eq10a}). For all distances the average energy of the $\Psi_2$ state is larger than the energy of the wavefunction of equation (\ref{eq10a}). However, we can achieve a lower average energy for the hydrogen molecule by describing the molecule by a wavefunction which is a linear combination of the two previously mentioned wavefunctions. This is the well known configuration interaction (CI) method \cite{atkins}.  Hence, in the CI method the wavefunctions is described by the following expression
\begin{equation}
\Psi=c_1 \Psi_1 +c_2 \Psi_2
\label{eq12}
\end{equation}
with $|c_1|^2 +|c_2|^2 =1$.

Using the wavefunction given by equation (\ref{eq12}) we can find lower values for the energy of the ground state of the hydrogen molecule. The energy of the $H_{2}$ molecule is for all distances lower than the energy of the wavefunction $\Psi_1$ as it is indicated in Fig. \ref{fig1} (compare the dashed-dotted  to the solid curve). Its minimum is at approximately $-0.237$ Hartree ($\approx -6.457$ eV) for $R \approx 1.67\alpha_0$. 

At this paragraph we show how in the context of the linear combination state (CI method) we estimate the coefficients $c_1$ and $c_2$ in order to achieve the lowest energy for any given internuclear distance $R$. The expectation value of energy is given by $E=|c_1|^2 H_{11}+c_1c_2^{\ast}H_{12}+c_1^{\ast}c_2H_{21}+|c_2|^2 H_{22}$, where $H_{11}(=E_{1})$ is the energy given by eq.(\ref{eq11}) and 
\begin{subequations}
\begin{align}
H_{12} &= \frac{m-j}{2(1-S^2)},\\
H_{21} &= \frac{m-j}{2(1-S^2)},\\
H_{22} &= 2E_{1s}+\frac{j_0}{R}-\frac{2j'-2k'}{1+S}+\frac{j+2k+m-4l}{2(1+S)^2}
\end{align}
\label{eq13a}
\end{subequations}
where we point out that the $H_{ij}$ are functions of the distance $s$. Now as $|c_1|^2 + |c_2|^2 =1$, we can set $c_1=\cos\omega$ and $c_2=\sin\omega$, assuming only real coefficients. The minimization procedure is taken by the vanishing of the derivative $\partial E/\partial \omega =0$. Consequently, we take the values of $c_1$ and $c_2$ for which the energy is least in every given distance $R$. More specifically, the coefficients $c_1$ and $c_2$ are given by the following equations

\begin{subequations}
\begin{align}
c_1=\frac{1}{2}+\frac{1}{2\sqrt{1+(\frac{2H_{12}}{H_{11}-H_{22}})^2}}\label{eq14a}\\
c_2=\frac{1}{2}-\frac{1}{2\sqrt{1+(\frac{2H_{12}}{H_{11}-H_{22}})^2}}\label{eq14b}
\end{align}
\end{subequations}

\begin{figure}[htbp!]
\centering
\includegraphics*[width=9 cm]{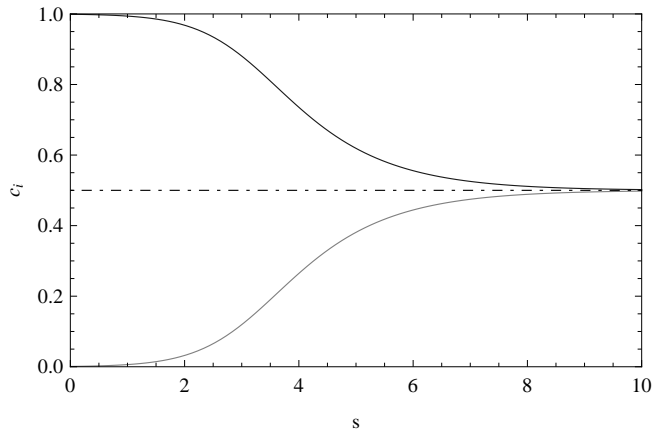}
\caption{Plots of the squares of the coefficients $c_i$($c_i^2$) for $i=1,2$ for which a minimum average energy is achieved as a function of distance $s=R/\alpha_0$ between the two protons in the $H_{2}$ molecule. The upper (lower) curve describes the $c_1$ ($c_2$) coefficient.}
\label{fig2}
\end{figure}

 Figure \ref{fig2} shows the dependence of the values of the squares of the $c_i$'s as a function of the internuclear distance. Note that the minimization predict negative values for the $c_2$.

Now, we turn our attention to the quantum entanglement that the two electrons may have in the ground state. As it was mentioned above, the wavefunction of two fermions is given by the eq. (\ref{eq1}). Also the single-particle space is four-dimensional and more specifically $\left\{|1\rangle, |2\rangle, |3\rangle, |4\rangle \right\}$. However, in our case we have to set the elements $w_{13}$ and $w_{24}$ of antisymmetric matrix equal to zero because in the ground state the total spin of electrons must be zero. As a result the antisymmetric coefficient matrix $w$ has the form
\begin{equation}
\left[ \begin{array}{cccc} 0 & w_{12} & 0 & w_{14} \\ -w_{12} & 0 & w_{23} & 0 \\ 0 & -w_{23} & 0 & w_{34} \\-w_{14} & 0 & -w_{34} & 0 \end{array} \right]
\label{eq13}
\end{equation}
Additionally, concurrence from eq. (\ref{eq6}) has the form
\begin{equation}
C(|w\rangle)=8|w_{12}w_{34}+w_{14}w_{23}| 
\label{eq14}
\end{equation}
By writing again the wavefunction for the ground state of the hydrogen molecule from eq. (\ref{eq12}) we assign the coeffients of each state with the corresponding ones from matrix of eq. (\ref{eq13}) and we find that 
\begin{subequations}
\begin{align}
w_{12}=w_{34}= \frac{c_1 c_+^2 + c_2 c_-^2}{2},\\
w_{14}=-w_{23}= \frac{c_1 c_+^2 + c_2 c_-^2}{2}
\end{align}
\label{eq15}
\end{subequations}
Substituting these values in eq. (\ref{eq14}) we take
\begin{equation}
C(|w\rangle)=2|c_1c_2|
\label{eq16}
\end{equation}

\begin{figure}[htbp!]
\centering
\includegraphics*[width=9 cm]{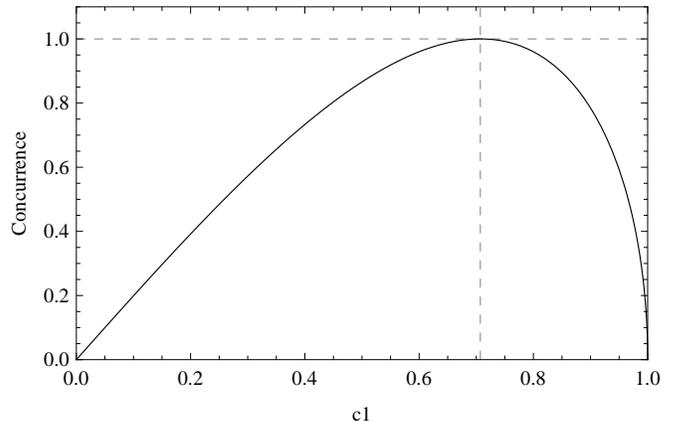}
\caption{Concurrence as a function of the coefficient $c_1$ for the ground state of the hydrogen molecule in the CI method.}
\label{fig3}
\end{figure}

\begin{figure}[htbp!]
\centering
\includegraphics*[width=9 cm]{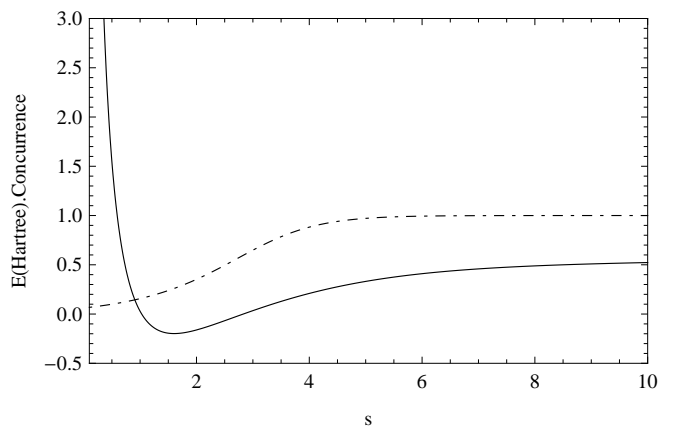}
\caption{This figure shows Con (dashed curve) and energy (solid curve) as a function of the internuclear distance. The energy have been calculated by CI method.}
\label{fig4}
\end{figure}

Figure \ref{fig3} shows how Con varies for different values of $c_1$. It is obvious that Con takes its maximum value for $c_1=\frac{1}{\sqrt{2}}$. However, from  Figs. \ref{fig1} and  \ref{fig2} we see that energy takes its minimum value for a different value of $c_1$. The  $c_1=\frac{1}{\sqrt{2}}$ corresponds to internuclear distances $R>$ larger than $8\alpha_0$ (i.e rather large distances). This fact is, also, amply demonstrated in Figure \ref{fig4} as we can see how the energy and Con vary as a function of the internuclear distance. It is evident that Con takes small values for small internuclear distances and becomes bigger as the distance R is rising. Finally, it becomes essentially unit for $R>8\alpha_0$.The internuclear distance for which the molecule has its minimum energy, i.e. has its most stable form, is $R \approx 1.67\alpha_0$. For this distance Con takes the value 0.2378. 

As a result, one may conclude that quantum entanglement is uncorrelated concerning the construction of a chemical bond. 

Furthermore, we know from \cite{schliemann3} that the product of elements $w_{12}w_{34}$ represent the entanglement which arises due to the orbital degrees of freedom and the product $w_{14}w_{23}$ the corresponding entanglement due to the spin degrees of freedom. It is easily seemed that in our case both kinds of entanglement contribute to concurrence.

To be more specific the fact the Con does not take its maximum value at the most stable state of the hydrogen molecule, most probably means that it is not the proper measure of the quantum correlations. Actually, this is what is now believed by the scientific community. Quantum discord provides all the quantum correlations and not simply the Con \cite{QD1,QD2}. This could be one of the future plans, as for the moment these is no straightforward definition for the quantum discord for indistinguishable particles.
Another future project, could be to study the case of finite temperature, where not only the ground state is occupied. In this case we believe a more obvious measure of the quantum correlations will be the entropy (see section II), as we now have a statistical mixtures. Hence, the chemical bond will correspond to maximization of a physical quantity that depend on both the energy and the entropy (i.e. the free energy of the system).  

\section{Conclusions}

In conclusion, we have presented a review of the theory for indistinguishable particles and we have reported two measures of quantum correlations, the concurrence and the von Neumann entropy. Next, taking advantage of this theoretical knowledge we implemented it in the prototype two- electrons hydrogen molecule. With the assistance of two well known approximations in molcecular physics (LCAO and CI methods) we systematically investigated the ground state of the $H_{2}$ molecule. As a result, we managed to find a closed form expression which estimates a typical measure of quantum entanglement (concurrence) for the two electrons when the molecule has its lowest energy. 

Finally, we expect (mainly by physical intuition) that the ground state of the molecule should be an entangled one. However, we have shown that the concurrence (an extensively used measure of quantum entanglement), is unrelated to the construction of the chemical bond. 
In this paper we computed the entanglement of the electrons which are in the hydrogen molecule as this is the simplest two- electrons system in the universe and the most well studied as well as we need the minimum number approximations in order to find its ground state wavefunctions. In addition, we believe that the present methodology and systematic theoretical research can be also applied to other much more complicated molecular systems.

\section*{Appendix}

Here we give some extra information concerning the integrals which we used in our computations. Again, we have in all of the following equations that $j_0=\frac{e^2}{4\pi \epsilon_0}$ as well as $s=R/\alpha_0$, with $\alpha_0$ the Bohr radius. Additionally, $\phi_\alpha$ and $\phi_\beta$ are the atomic orbitals of each hydrogen atom in the ground state. The required integrals for all computations are:
\begin{equation}
\frac{j'}{j_0}=\int \frac{\phi_\alpha^2(1)}{r_{1b}}d\tau_1=\frac{1}{R}[1-(1+s)e^{-2s}]
\label{eq17}
\end{equation}
\begin{equation}
\frac{k'}{j_0}=\int \frac{\phi_{\alpha}(1)\phi_{\beta}(1)}{r_{1b}}d\tau_1=\frac{1}{\alpha_0}(1+s)e^{-s}
\label{eq18}
\end{equation}
\begin{equation}
\frac{j}{j_0}=\int \frac{\phi_{\alpha}^2(1)\phi_{\beta}^2(2)}{r_{12}}d\tau_1d\tau_2=\frac{1}{R}-\frac{1}{2\alpha_0}(\frac{2}{s}+\frac{11}{4}+\frac{3}{2}s+\frac{1}{3}s^2)e^{-2s}
\label{eq19}
\end{equation}
\begin{equation}
\frac{k}{j_0}=\int \frac{\phi_{\alpha}(1)\phi_{\beta}(1)\phi_{\alpha}(2)\phi_{\beta}(2)}{r_{12}}d\tau_1d\tau_2=\frac{A(s)-B(s)}{5\alpha_0}
\label{eq20}
\end{equation}
\begin{equation}
\frac{l}{j_0}=\int \frac{\phi_{\alpha}^2(1)\phi_{\alpha}(2)\phi_{\beta}(2)}{r_{12}}d\tau_1d\tau_2=\frac{1}{2\alpha_0}[(2s+\frac{1}{4}+\frac{5}{8s})e^{-s}-(\frac{1}{4}+\frac{5}{8s})e^{-3s}]
\label{eq21}
\end{equation}
\begin{equation}
\frac{m}{j_0}=\int \frac{\phi_{\alpha}^2(1)\phi_{\alpha}^2(2)}{r_{12}}d\tau_1d\tau_2=\frac{5}{8\alpha_0}
\label{eq22}
\end{equation}
with
\begin{equation}
A(s)=\frac{6}{s}[(\gamma+\ln{s})S^2-E_1(4s)S'^2+2E_1(2s)SS']
\label{eq23}
\end{equation}
\begin{equation}
B(s)=[-\frac{25}{8}+\frac{23}{4}s+3s^2+\frac{1}{3}s^3]e^{-2s}
\label{eq24}
\end{equation}
\begin{equation}
S(s)=(1+s+\frac{1}{3}s^2)e^{-s}
\label{eq25}
\end{equation}
\begin{equation}
S'(s)=S(-s)=(1-s+\frac{1}{3}s^2)e^{s}
\label{eq26}
\end{equation}
 Finally $\gamma$ is Euler's constant and $E_1(x)$ is the well-known exponential integral which is given by the equation
\begin{equation}
E_1(x)=\int_{x}^{\infty} \frac{e^{-z}}{z}dz
\label{eq26}
\end{equation}

{}

\end{document}